# On the Radiation Damage of BTeV EMCAL Detector Unit[1]

A. Borisevich, V. Dormenev, G. Drobychev, A. Fyodorov, M. Korzhik, Y. Kubota*, A. Lobko, A. Lopatik, O. Missevitch, V. Moroz

Belarussian State University, Institute for Nuclear Problems, Minsk, BELARUS
* University of Minnesota, Minneapolis, USA

## Abstract

Hamamatsu photomultiplier tubes (PMT) and various PMT window materials were exposed to gamma irradiation. Tests were performed with absorbed doses of 1 krad and 120 krad. Initial recommendations on PMT types to use in the BTeV electromagnetic calorimeter are stated.

## 1. Introduction

The electromagnetic calorimeter (EMCAL) of the BTeV detector will consist of Lead Tungstate ($PbWO_4$, PWO) crystals and photomultiplier tubes. Radiation hardness of PWO crystals has been studied very carefully for many years (see Ref. [1], for example). Radiation hardness of PMTs, which will also be crucial elements of the calorimeter, still needs to be studied. Because there PMTs will detect scintillation light from the PWO crystals, reduction in the PMT window transparency in the UV region will crucially affect the performance of EMCAL. In addition, exposure to high-energy gamma-rays with energies more than 1 MeV will increase the dark current considerably, which worsen the noise performance [2]. Charged particles may damage photo-cathode, dynode system, etc. taking into account the specific radiation environment in which EMCAL will operates.

The PWO crystals located in front of PMTs will act as radiation converters absorbing much of the energies of high-energy gamma rays, electrons and hadrons while producing many soft photons and neutrons by interactions, PMTs will be exposed to these particles in addition to primary hadrons, photons and electrons. Much work has been done to estimate the dose rate in the BTeV EMCAL environment [3] including flux of neutrons and very soft photons. Given the results of this study, we will need to study PMT radiation hardness using various types of radiation and various dose rates.

The relationship between absorbed energy of radiation and radiation-induced effects is often not proportional but rather complex. In addition to the total radiation dose we often need to know the time dependence of the radiation, and the types and energies of the ionizing radiation.

As a first step to study the issue experimentally we used a $^{60}Co$ gamma ray source ($E$~1.25 MeV) to evaluate radiation-induced damage of PMTs. For a glass such as $SiO_2$ the Compton effect will dominate in interaction for low-energy (below 1 MeV) gamma rays. Given the typical thickness of PMT windows of 1-2 mm, the absorbed dose caused by soft gamma-quanta will be larger than the one caused by 1.25-MeV gamma rays at the same exposure dose, which is usually easier to evaluate. In other words, test with 1.25 MeV gamma-quanta gives us a lower limit on the effect arising from soft gamma rays. Dose values in our current study were chosen to be the same order of magnitude as they were calculated for the BTeV detector [3].

## 2. Experiments and Results

### 2.1 Radiation damage studies of R5800 Hamamatsu PMT

#### 2.1.1 Investigated samples and experimental conditions

Two R5800 Hamamatsu PMTs (serial numbers CA0622 and CA0641) were used in beam tests prior to our studies but there was no indication of irreversible effects caused by previous irradiation for either PMT. We have used one of them (#CA0641) for our irradiation tests and the other PMT was used as a control sample. Stability of data acquisition system was periodically checked and was found to be as good as 1%.

[1] This work was supported in part by NATO PST Collaborative Linkage Grant “Study of PWO Crystals for BTeV”.

Irradiation was performed with a $^{60}$Co source ($E_\gamma$~1.25 MeV) whose activity provided about 100 R/s exposure dose rate. Time of irradiation was ~15 minutes, so absorbed dose was estimated to be about 120 krad. Only 3.5% of BTeV PMT's will receive this much absorbed dose in a year. Measurements of PMT properties were started in approximately 40 minutes after the completion of irradiation and took about five minutes.

Changes in the PMT parameters have been evaluated through pulse height measurements at four wavelengths, two of which were generated by scintillator-based sources and two were produced by LED-based sources. Their parameters are listed in Table 1 below.

Table 1.

| Light Source | Wavelength | | Reference |
|---|---|---|---|
| | Peak | FWHM | |
| (YAP:Ce–$^{241}$Am) Light Pulser Source (LPS) | 360 nm | 60 nm | [4] |
| NaI(Tl) 25∅*1.0 mm – $^{137}$Cs | 415 nm | 120 nm | [3] |
| LED Pulser-1 | 515 nm | 70 nm | [5, 6] |
| LED Pulser-2 | 660 nm | 25 nm | |

### 2.1.2 Pulse-height spectrum measurements

Examples of pulse-height spectra measured using LPS with both PMTs at the same high voltage before and after irradiation are shown in Fig. 1.

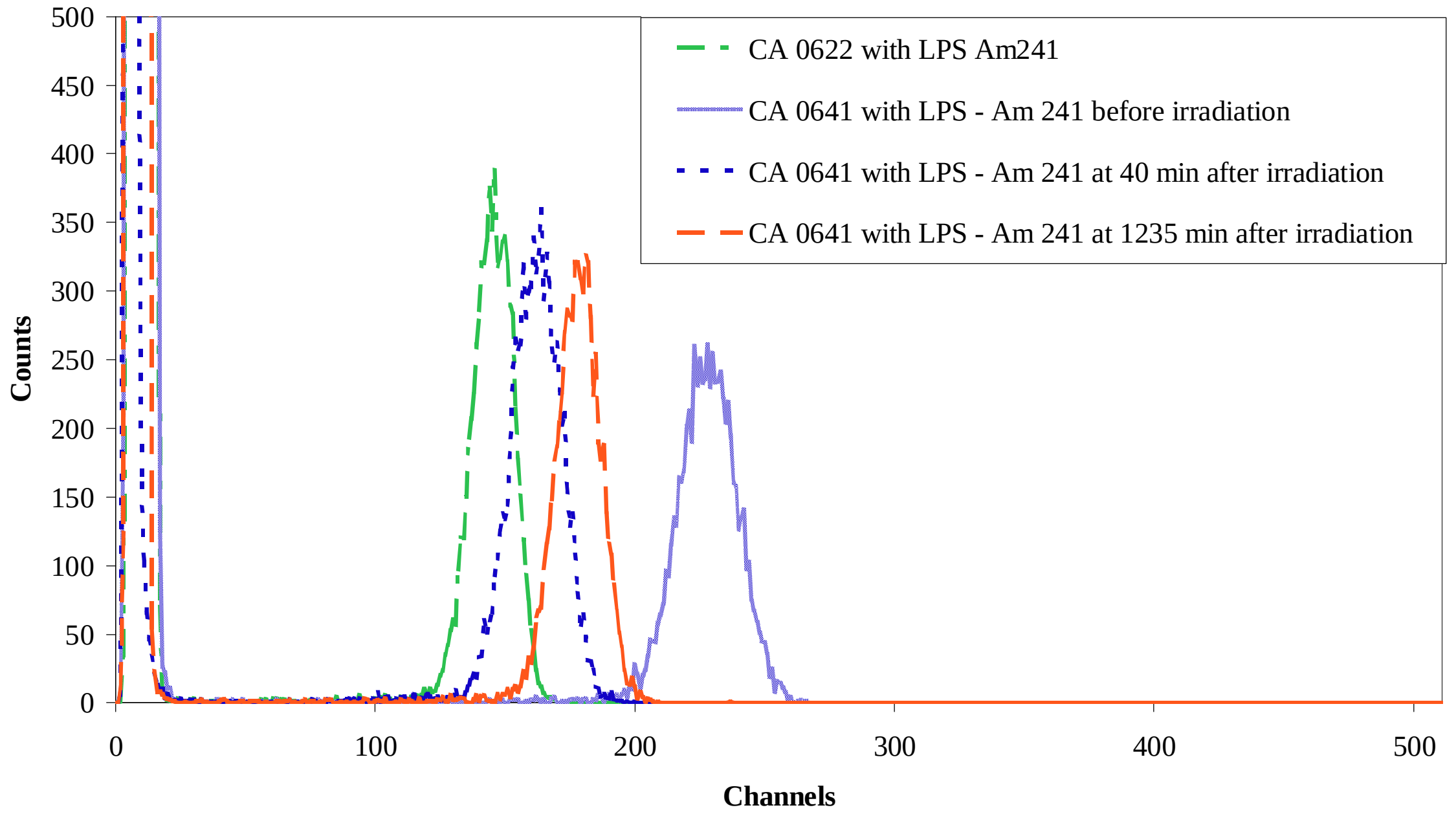

Fig. 1. Pulse height spectra measured with LPS. Since the pulse height spectra for CA 0622 (control sample) didn't change throughout the study we show only one distribution for clarity.

In Fig. 2 we have summarized relative changes of amplitudes of the signals produced by various light sources after irradiation. One can see here that R5800 PMT has the largest deterioration in ultraviolet range.

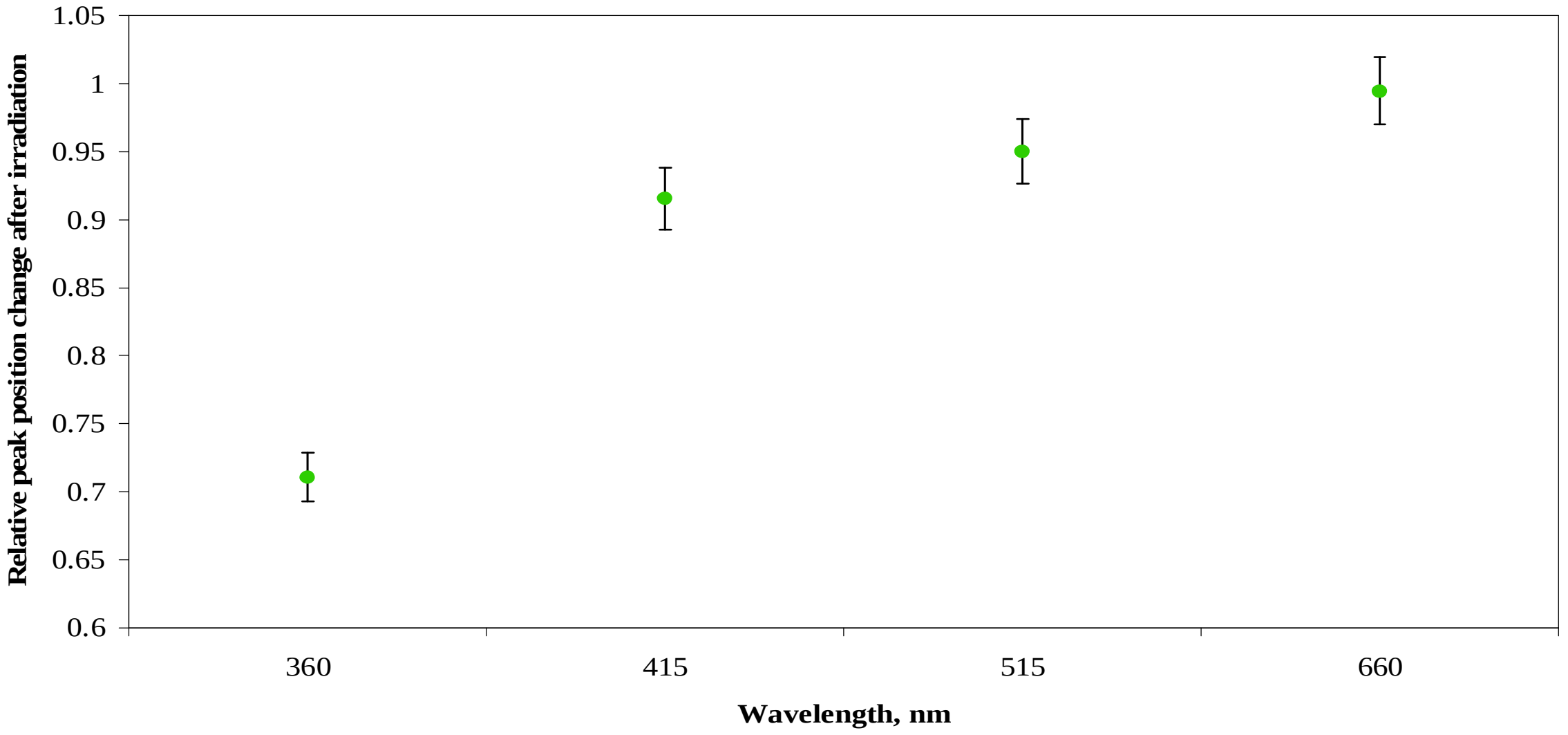

Fig. 2. Change of relative peak positions after R5800 PMT irradiation.

The PMT recovery at 360 nm is shown in Fig. 3. We estimate that the recovery time constant is not less than 250 hours. As R5800 PMT have reached full (>95%) recovery, we should conclude that applied doses do not produce irreversible damages with $^{60}Co$ irradiation of 120 krad.

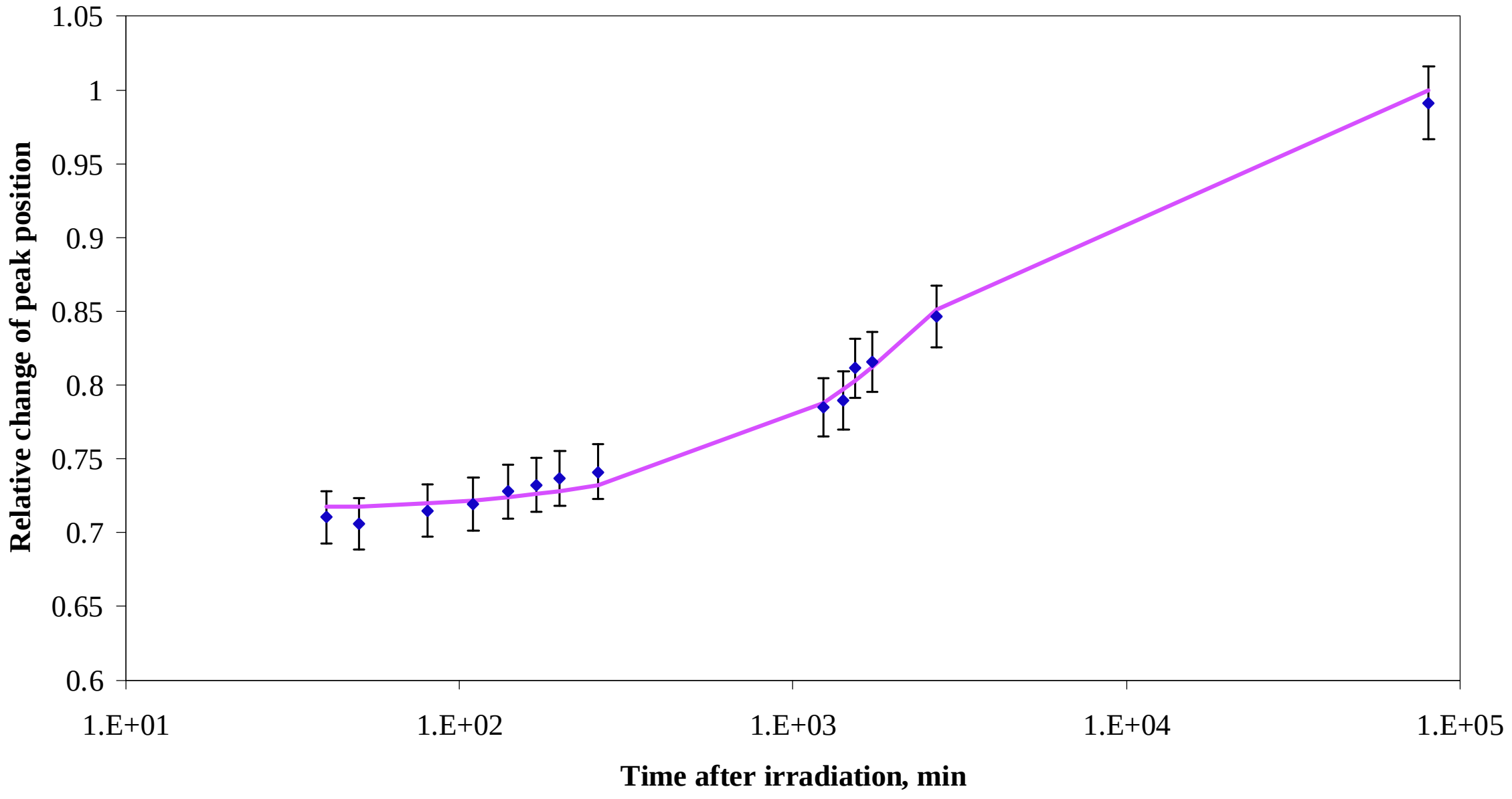

Fig. 3. Recovery of R5800 PMT at 360 nm measured by the change of the LPS peak position.

### 2.1.3 Integral change of PMT sensitivity as result of radiation damage

For additional verification of the results described above, we have used another technique and studied change of PMT sensitivity spectra measured for R5800 Hamamatsu PMT (#CA0641) after irradiation using modulated light from monochromator of luminescent spectrometer as a source. Thus, we have measured PMT sensitivity using monochromatic light source at wavelengths 340, 360, 380, 400, 410, 420, 430, 440, 460, 480, 500, 520, 550 and 600 nm. Fig. 4 shows PMT sensitivity spectra measured one hour after the 120-krad irradiation and one day afterward normalized on the spectrum measured before the irradiation. For comparison, typical PWO emission spectrum is also shown. We can clearly see that PMT radiation-induced damage wavelength range overlaps to a great extent with the PWO emission band.

In order to estimate the net effect of the PMT deterioration on the PWO scintillation light detection, PMT sensitivity spectra were multiplied by PWO emission spectrum and then integrals over the wavelength were calculated. The values of these integrals are presented in Table 2. Since a loss of 30% in the light signal is not acceptable for BTeV, R5800 PMT or, generally, a PMT equipped with a borosilicate glass window cannot be used in the BTeV EMCAL.

Table 2.

| Description | Relative value of integral |
|---|---|
| Before irradiation | 1.00 |
| 1 hour after irradiation | 0.69 |
| 1 day after irradiation | 0.77 |

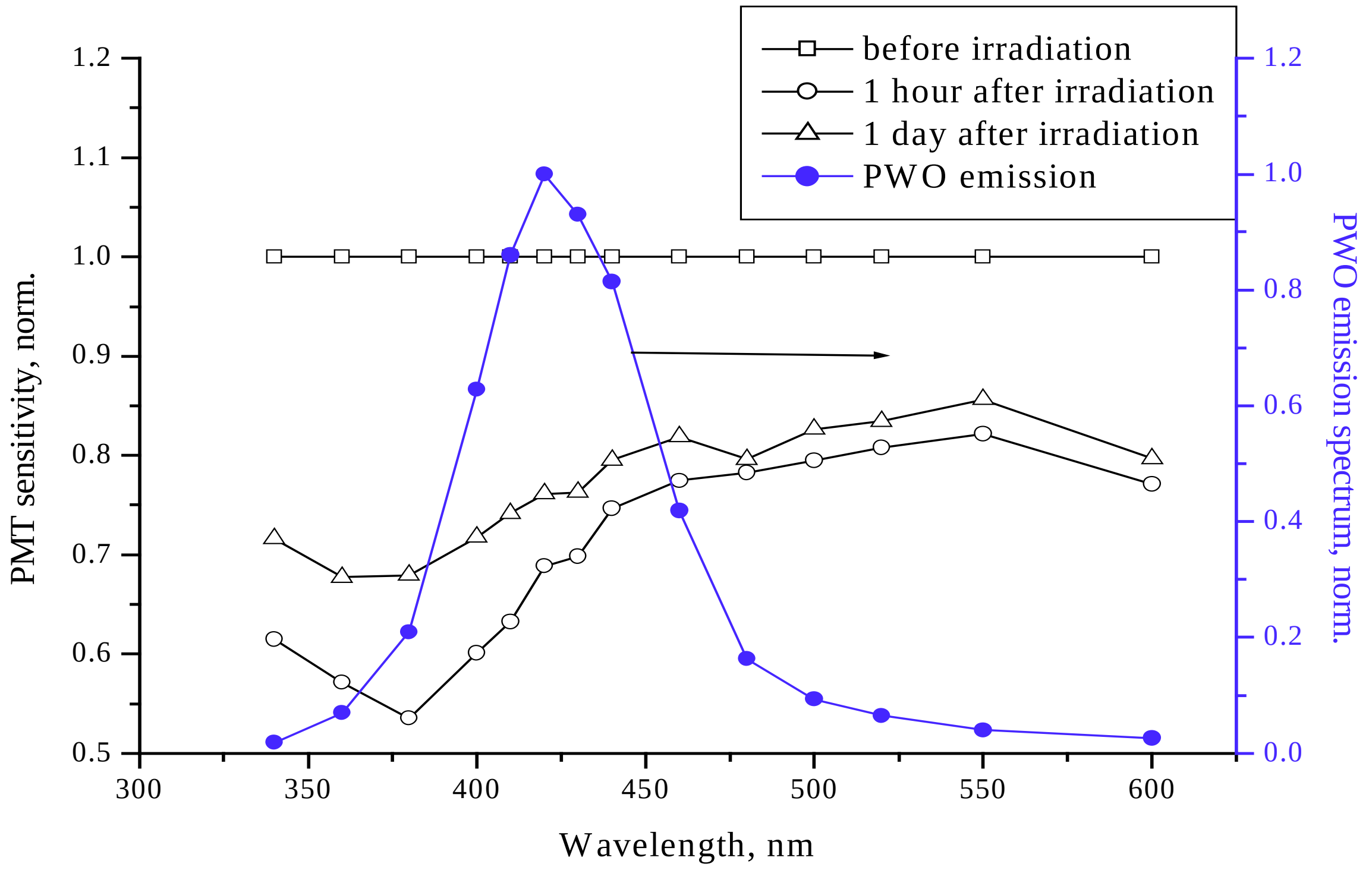

Fig. 4. PMT sensitivity spectra measured one hour after irradiation (open circles)/one day after irradiation (open triangles), and normalized by the spectrum measured before irradiation. Filled circles show the PWO emission spectrum.

## 2.2 Study of radiation damage of the PMT window materials

To find more suitable PMT window materials for BTeV applications, we have tested six samples of PMT window materials with dimensions 25.0∅×1.0 mm sent by Hamamatsu Corporation. They were made of three different materials as listed below:

1. Borosilicate glass - 2 pieces;
2. UV glass - 2 pieces;
3. quartz - 2 pieces.

We also included two pieces made of industrial sapphire ($Al_2O_3$) microchip substrate, which has been suggested to be good UV transparent PMT window material [7], in shape of parallelepiped with dimensions 25.0×25.0×0.4 $mm^3$.

120 krad tests were performed when one sample of each pair was irradiated for 15 minutes by gamma-rays from a $^{60}Co$ source with non-adjustable exposure dose rate of about 100 R/s. Resulting induced absorption spectra are presented in Fig. 5a-d. For comparison, typical radiation-induced absorption spectrum of CMS-type PWO crystal at the same irradiation conditions is shown in Fig. 6.

Recovery of radiation-induced absorption was measured for Borosilicate and UV glasses. Results drawn for 460 nm (that is the wavelength of the LED-based reference light source recently designed as a prototype for the BTeV EMCAL monitoring system) are presented in Fig. 7a and b.

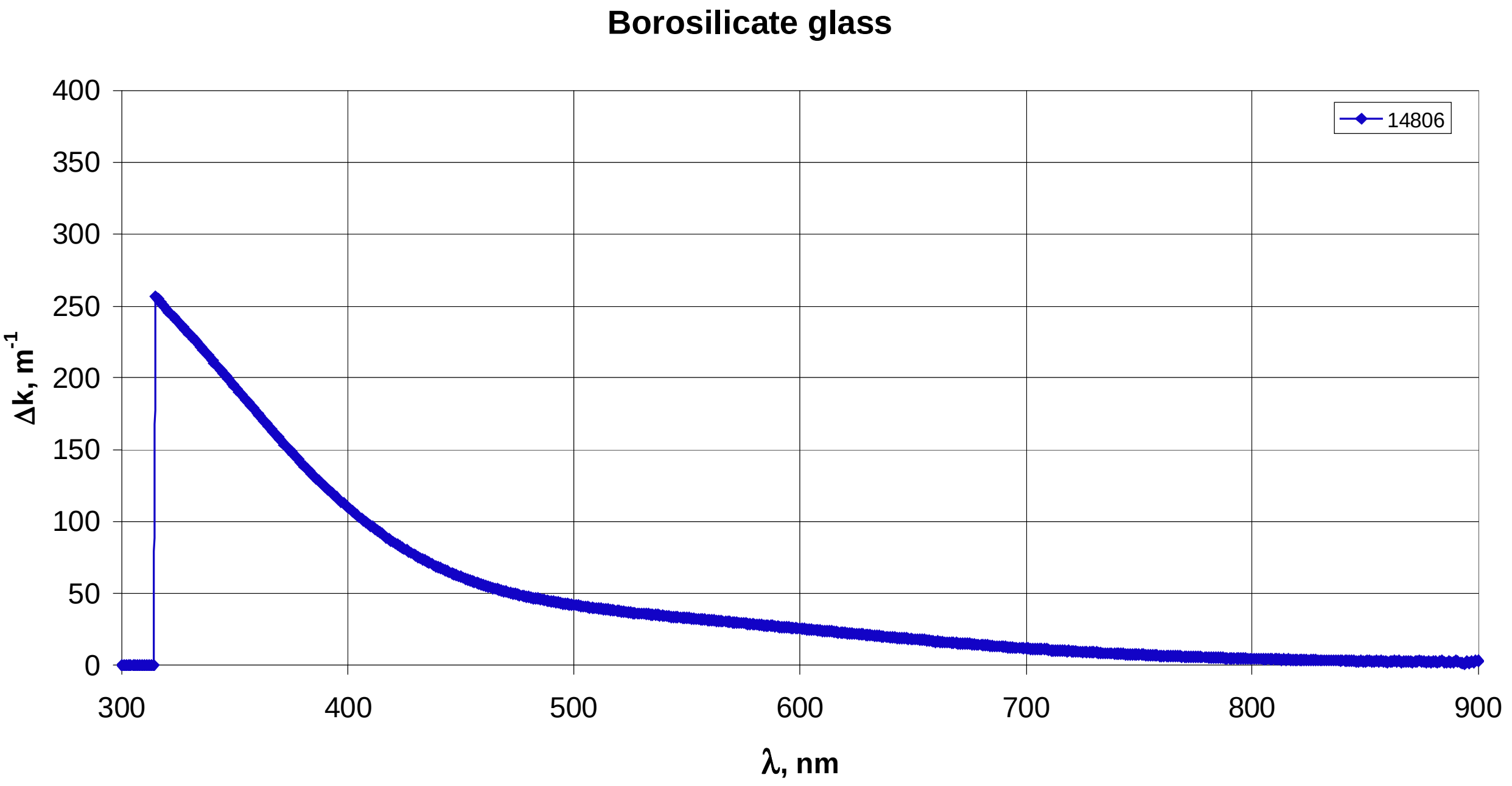

Fig. 5a Radiation-induced absorption of the Borosilicate glass sample after 120-krad irradiation

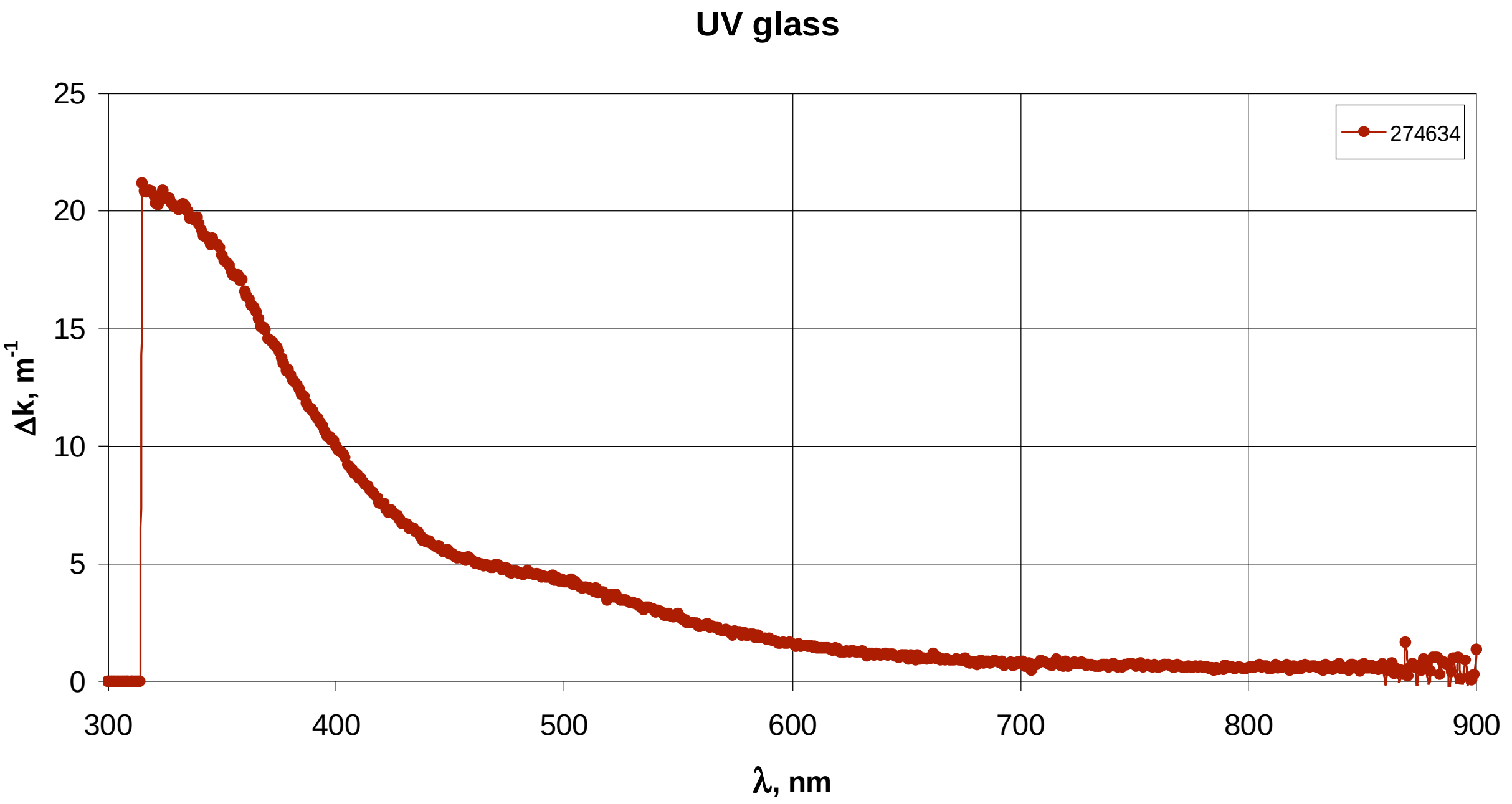

Fig. 5b Radiation-induced absorption of the UV glass sample after 120-krad irradiation

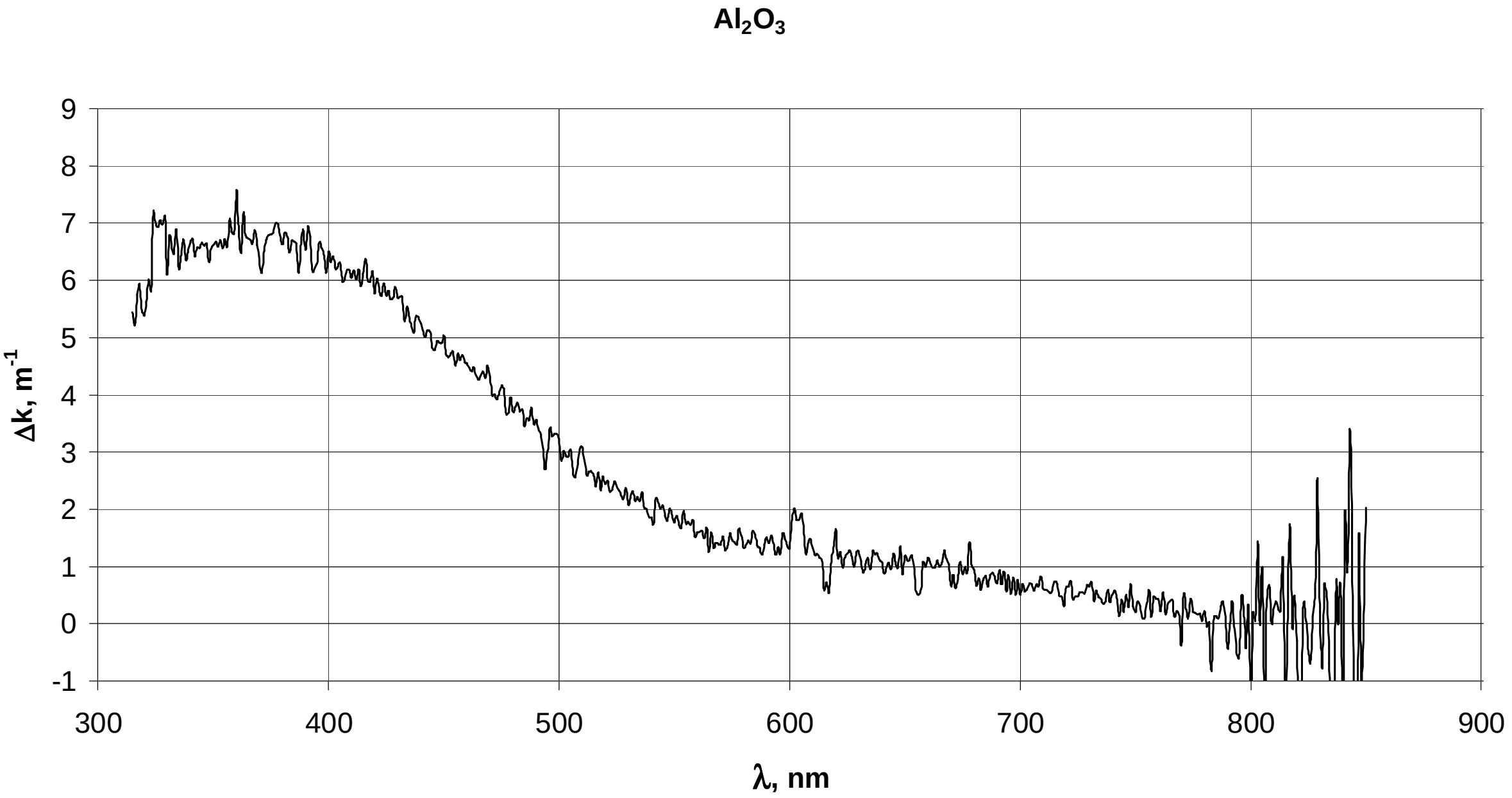

Fig. 5c Radiation-induced absorption of the sapphire sample after 120-krad irradiation

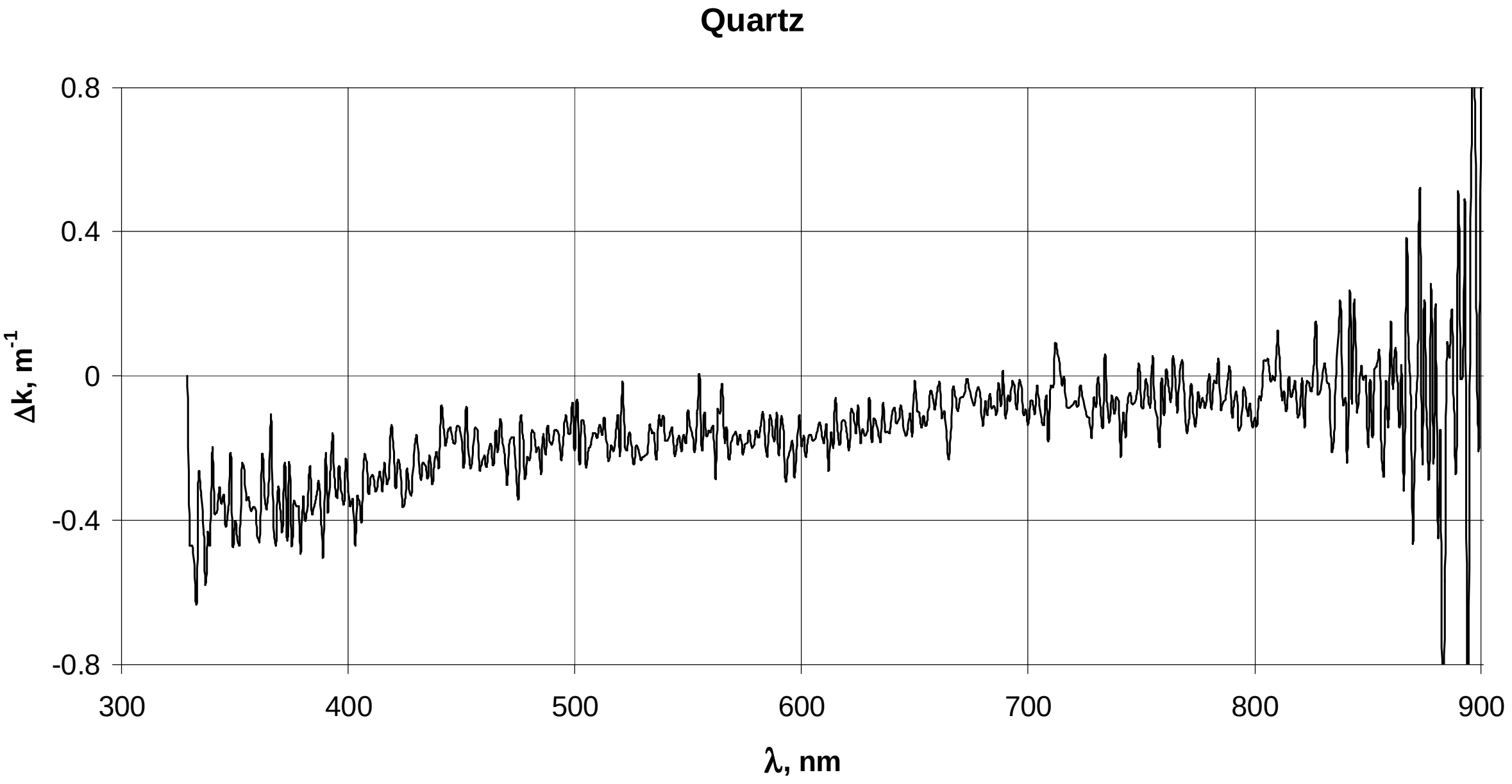

Fig. 5d Radiation-induced absorption of quartz sample after 120-krad irradiation

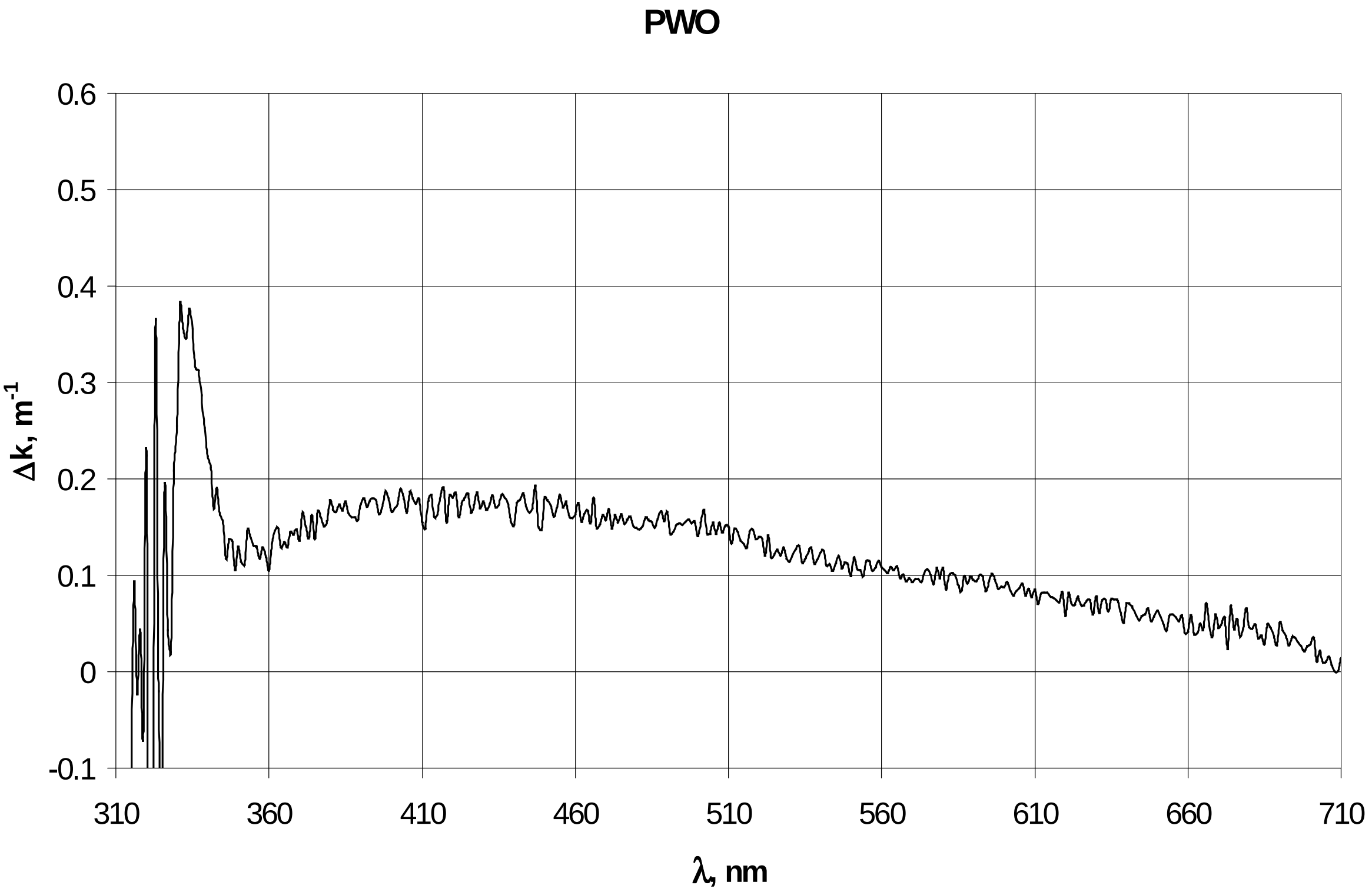

Fig. 6 Radiation-induced absorption of the typical PWO crystal after 120-krad irradiation

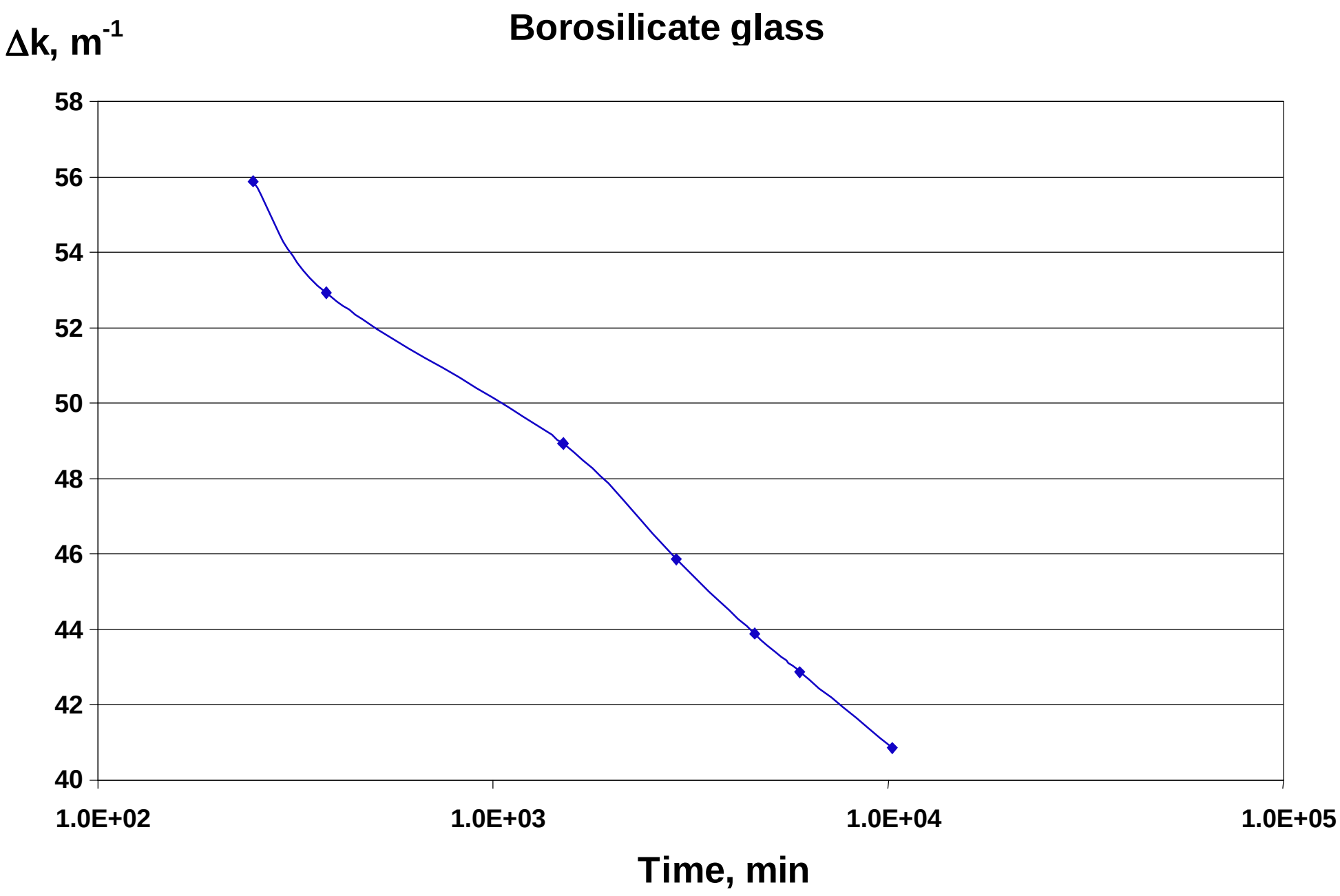

Fig. 7a Recovery of Borosilicate glass induced absorption at 460 nm

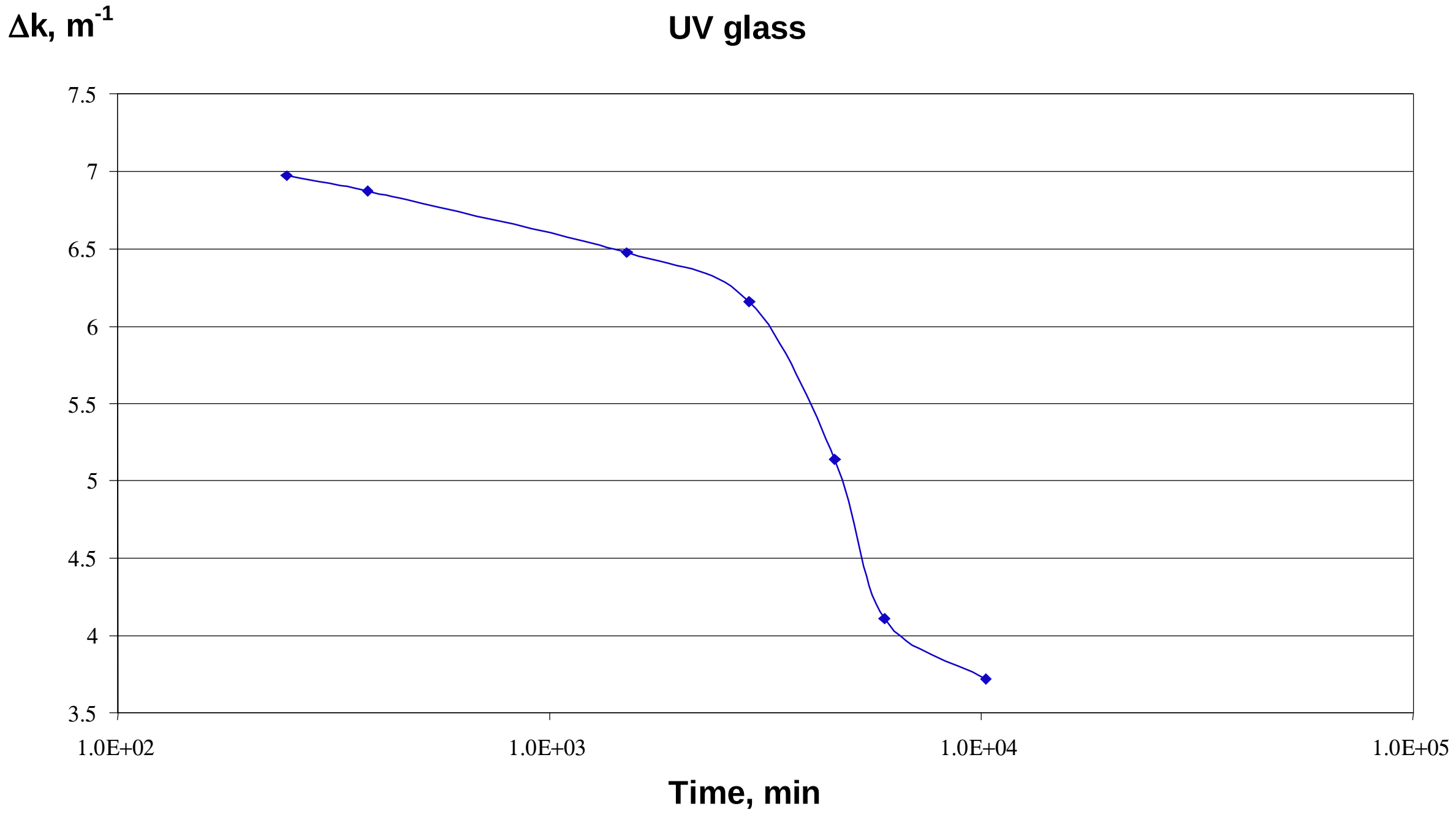

Fig. 7b Recovery of UV glass induced absorption at 460 nm

Finally, to compare with our results, data from the Hamamatsu Handbook [8] have been used to calculate the induced absorption *vs.* exposure dose. These results are displayed in Fig. 8. Extrapolating this curve to small exposure doses, we estimate that the radiation-induced absorption in this material is about 50 $m^{-1}$ at $10^4$ R exposure dose which is consistent with what we observe with our Borosilicate glass window sample from Hamamatsu.

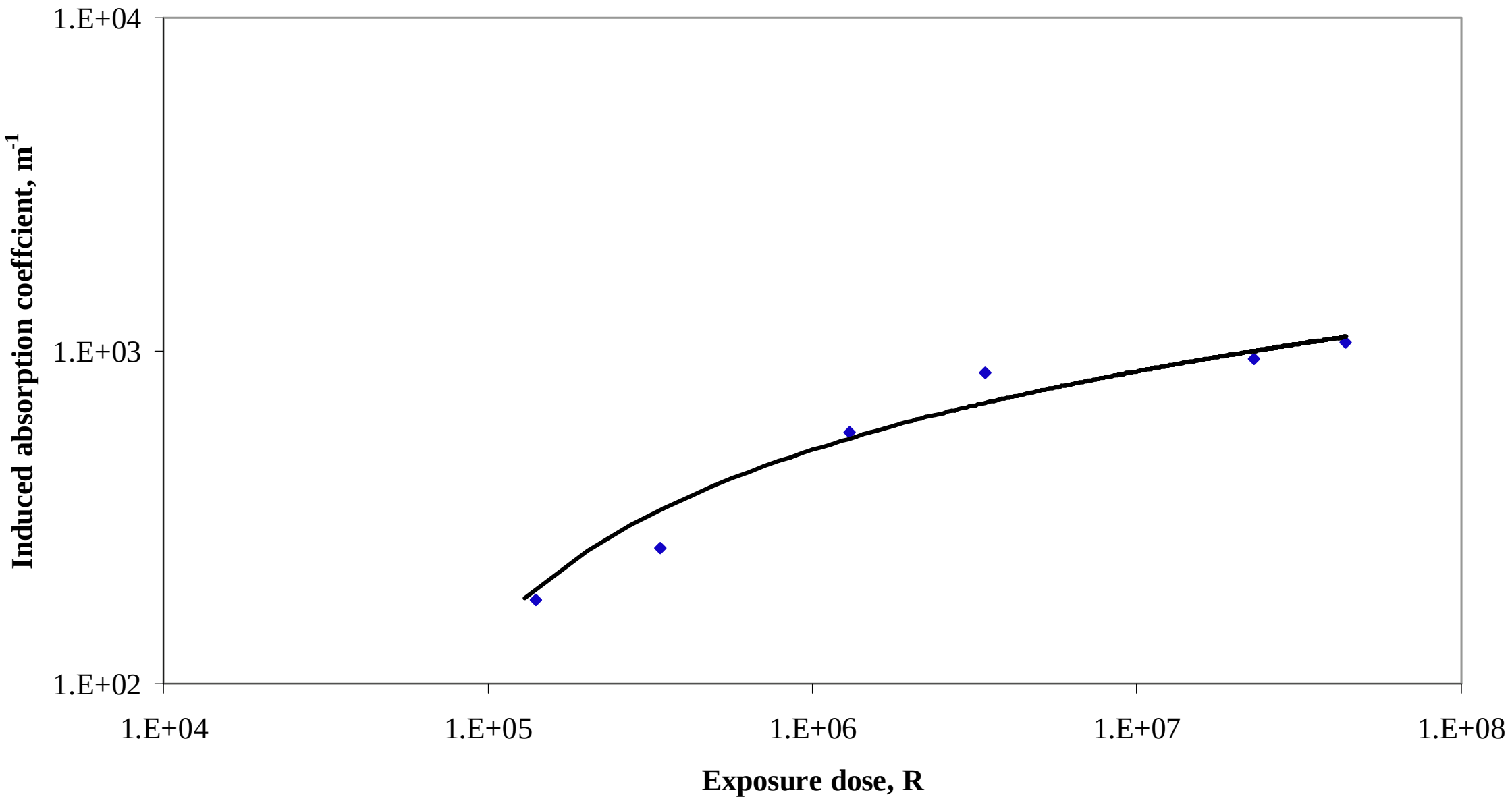

Fig. 8. Induced absorption at 400 nm for borosilicate glass (data extracted from [8]).

From these observations, we conclude that:

- The Borosilicate glass windows have significant damages (induced absorption of ~90 $m^{-1}$ at 420 nm) after 120 krad (~$10^3$ Gy) gamma-irradiation;

- The UV glass windows' damages are about 10 times lower than that of the Borosilicate glass under the same irradiation conditions, but remain still significantly larger (~8 $m^{-1}$ at 420 nm) than typical PWO radiation damage (~0.2 $m^{-1}$ at 420 nm);
- Sapphire sample damage have the same order of magnitude as that of the UV glass;
- No observable damages of the quartz sample were detected;
- The glass samples have slow (days) recovery after irradiation.

We also tested these samples using a lower and adjustable intensity $^{137}Cs$ source (E=0.662 MeV). The total dose was 1 krad.

Only two kinds of window materials were chosen for these low dose tests, namely Borosilicate and UV glasses, since the quartz window did not show any deterioration in the high dose tests. The results are presented in Fig. 9a,b and displayed in Table 3 together with previous results.

Table 3

| Material | Absorbed dose, krad | Induced absorption at 420 nm, $m^{-1}$ |
|---|---|---|
| Borosilicate glass | 1 | 1.0±0.1 |
| | 120 | 90±10 |
| UV glass | 1 | 0.4±0.1 |
| | 120 | 8.0±0.3 |

As one can see, the rate of damage generation in the Borosilicate glass is much higher than that of the UV glass.

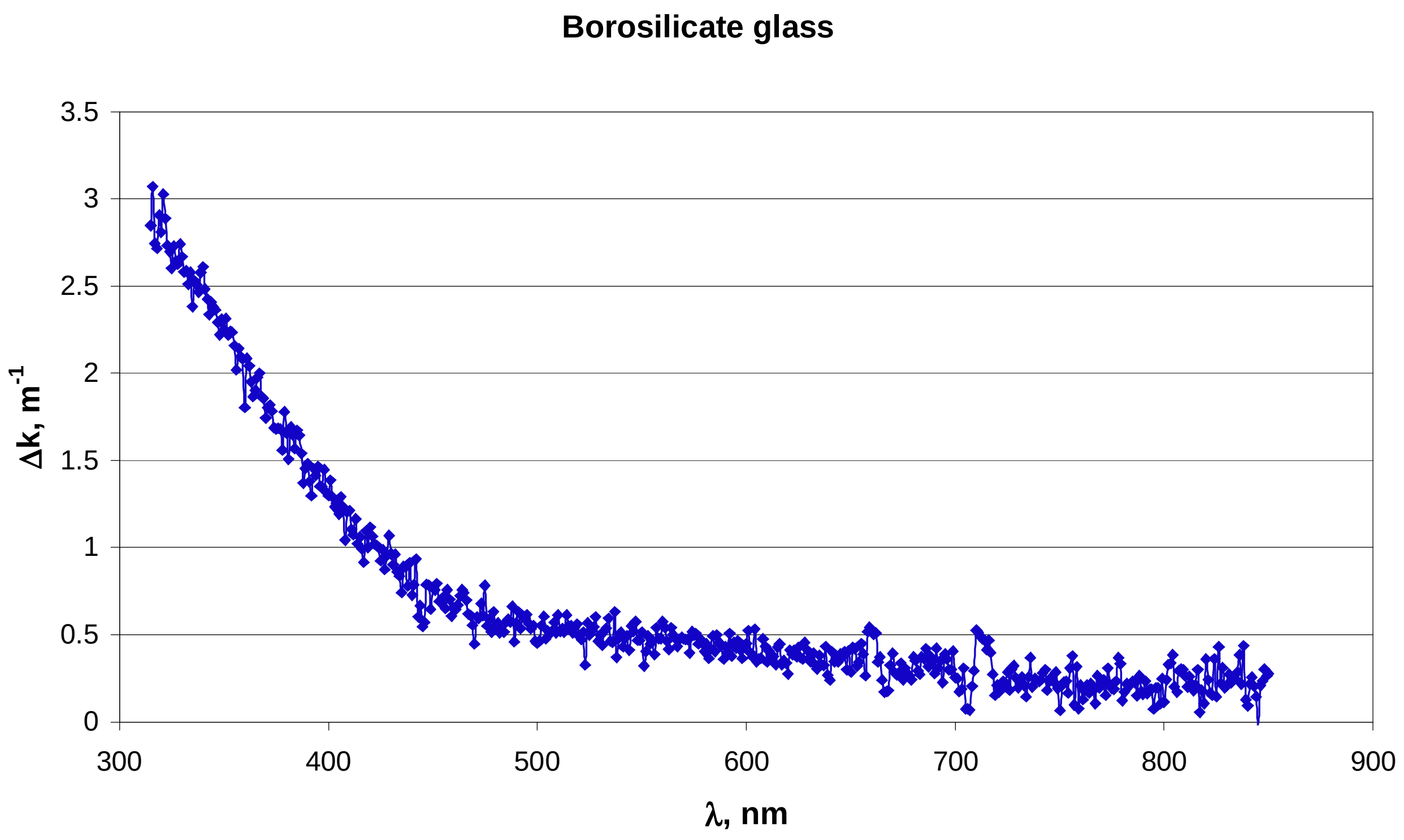

Fig. 9a Radiation-induced absorption of the Borosilicate glass sample after 1 krad irradiation

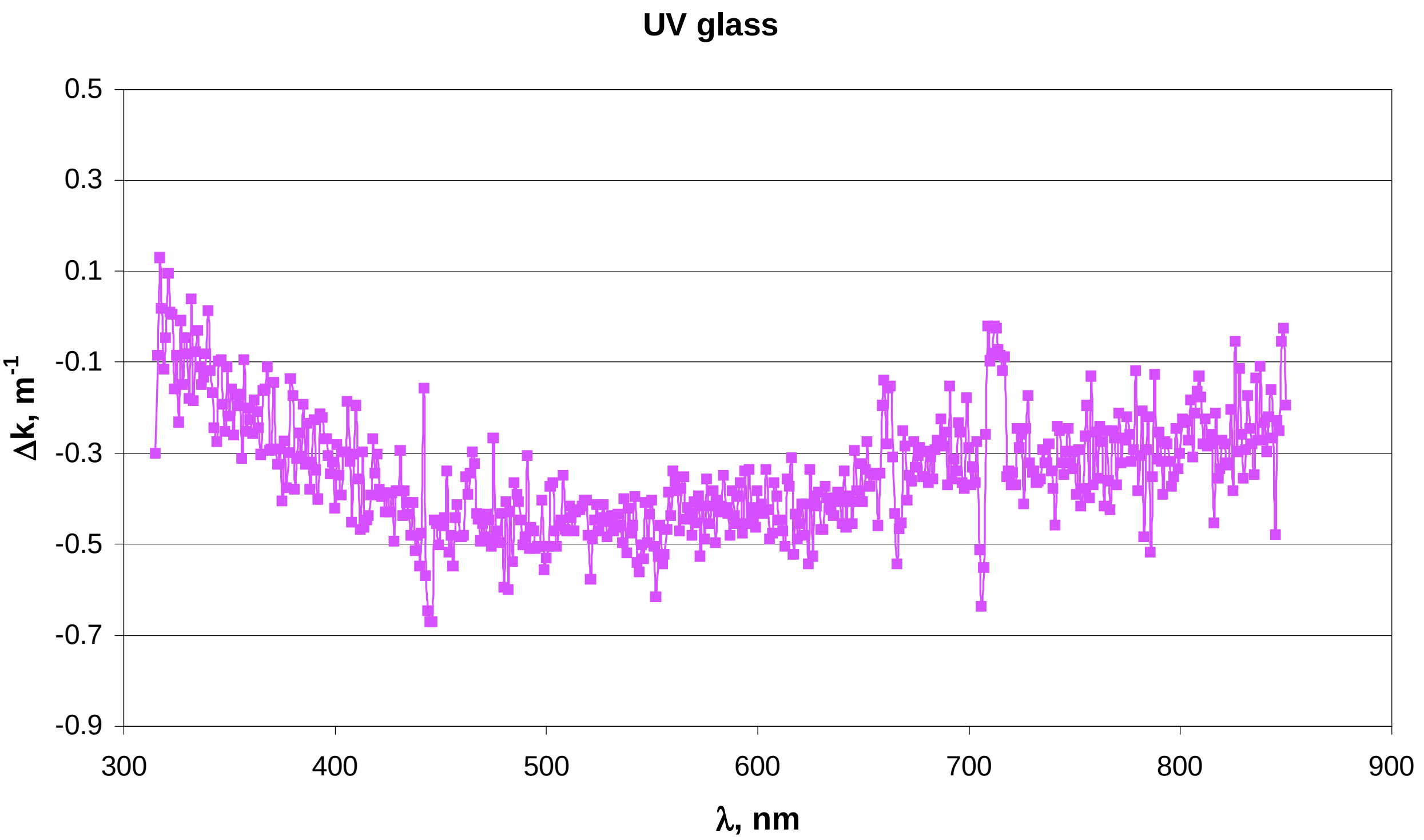

Fig. 9b Radiation-induced absorption of the UV glass sample after 1 krad irradiation

## 3. Conclusions

Comparative tests on gamma-irradiation of industrial photo-multipliers and various PMT window materials confirmed our assumption that in the relevant range of doses and dose rates, degradation of PMT sensitivity for PWO scintillation light depends only on PMT window transmission loss in UV range.

Negligible damage of the window optical transmission is observed in red range of visible light spectrum. Therefore, it can be used for additional and separate monitoring of readout stability of BTeV EMCAL PMT-based detector unit.

Taking into account radiation environment inside the BTeV calorimeter we may recommend considering a usage of PMT with quartz windows at least in its central part. On the other hand, it is very likely to use less expensive PMT equipped by UV glass windows in the periphery where the radiation levels are significantly lower.

Our results are similar to the studies of vacuum photo-triodes for the CMS end-caps [9], even though much higher doses and dose rates were used for the latter.

Finally, we must also mention again that our research was performed using gamma-ray sources, so the results are somewhat indirect. As before, it is important to understand PMT radiation hardness in fields of ionizing radiation produced by hadrons behind PWO crystals, which work as a radiation converter.